\documentclass{aa}  
\usepackage{color}
\usepackage{graphicx}
\usepackage{txfonts}
\usepackage{lipsum}
\usepackage{subcaption}        
\usepackage{lscape}                                         
\usepackage{placeins}           
\usepackage{threeparttable}
\usepackage{rotating}
\usepackage{multirow}
\usepackage{booktabs}  
\usepackage{natbib}
\bibpunct{(}{)}{;}{a}{}{,} 
\usepackage{hyperref}

\begin{document}

   \title{The MALATANG survey: star formation, dense gas, and AGN feedback in NGC~1068}

   \author{Shuting Lin\inst{1}
        \and Siyi Feng\inst{1} 
        \and Zhi-Yu~Zhang\inst{2,3} \and Chunyi Zhang\inst{1} \and Qing-Hua~Tan\inst{4} \and Junzhi~Wang\inst{5} \and Yu~Gao\inst{1} \and Xue-Jian~Jiang\inst{6} \and Yang~Gao\inst{7,4} \and Xiao-Long~Wang\inst{8,9,10} \and Junfeng Wang\inst{1} \and Jian-Fa Wang\inst{4} \and Satoki~Matsushita\inst{11} \and Aeree~Chung\inst{12} \and Kotaro~Kohno\inst{13} \and Tosaki~Tomoka\inst{14} \and Thomas R. Greve\inst{15,16,17}
        }

   \institute{Department of Astronomy, Xiamen University, Zengcuo'an West Road, Xiamen, 361005\\
             \email{syfeng@xmu.edu.cn}
            \and  School of Astronomy and Space Science, Nanjing University, Nanjing 210093 
            \and Key Laboratory of Modern Astronomy and Astrophysics (Nanjing University), Ministry of Education, Nanjing 210093
            \and Purple Mountain Observatory, Chinese Academy of Sciences, 10 Yuanhua road, Nanjing 210023 
            \and School of Physical Science and Technology, Guangxi University, Nanning 530004
            \and Research Center for Astronomical Computing, Zhejiang Laboratory, Hangzhou 311121
            \and Shandong  Key Laboratory of Space Environment and Exploration Technology, College of Physics and Electronic Information, Dezhou University, Dezhou 253023
            \and Department of Physics, Hebei Normal University, Shijiazhuang 050024 (Physics Postdoctoral Research Station at Hebei Normal University)
            \and Guo Shoujing Institute for Astronomy, Hebei Normal University, Shijiazhuang 050024
            \and Hebei Advanced Thin Films Laboratory, Shijiazhuang 050024
            \and Institute of Astronomy and Astrophysics, Academia Sinica, 11F of Astro-Math Building, AS/NTU, No.1, Sec. 4, Roosevelt Rd, Taipei 10617
            \and Department of Astronomy, Yonsei University, 50 Yonsei-ro, Seodaemun-gu, Seoul 03722, Republic of Korea
            \and Institute of Astronomy, Graduate School of Science The University of Tokyo 2-21-1 Osawa, Mitaka, Tokyo, 181-0015, Japan
            \and Joetsu University of Education, Yamayashiki-machi, Joetsu, Niigata 943-8512, Japan
            \and Cosmic Dawn Center (DAWN),  Copenhagen, Denmark
            \and DTU-Space, Technical University of Denmark, Elektrovej 327, DK-2800 Kgs. Lyngby, Denmark
            \and Department of Physics and Astronomy, University College London, Gower Street, London WC1E 6BT, UK
            }

   \date{Received October 21, 2025}

  \abstract
   {}
   {We aim to investigate the interplay between dense molecular gas, star formation, and active galactic nucleus (AGN) feedback in the luminous infrared galaxy (LIRG) NGC 1068 at sub-kiloparsec scales.} 
   {We present the HCN~(4-3) and HCO$^+$~(4-3) maps of NGC~1068, obtained with JCMT as part of the Mapping the dense molecular gas in the strongest star-forming galaxies (MALATANG) project, and perform spatially resolved analyses of their correlations with infrared luminosity and soft X-ray emission.}
   {Spatially resolved relations between the luminosities of infrared dust emission and dense molecular gas tracers ($L_{\rm IR}-L'_{\rm dense}$) are found to be nearly linear, without clear evidence of excess contributions from AGN activity. 
   The spatially resolved X-ray emission ($L^{\rm gas}_{0.5\text{–}2\,\mathrm{keV}}$) 
   displays a radially-dependent twofold correlation with the star formation rate (SFR), suggesting distinct gas-heating mechanisms between the galaxy center and the outer regions. A super-linear scaling is obtained in galactic center regions with SFR surface density ($\Sigma_{\rm SFR}$) $>$ 8.2 $\times$ 10$^{-6}$ $M_\odot$ yr$^{-1}$ kpc$^{-2}$: log($L^{\rm gas}_{0.5\text{–}2\,\mathrm{keV}}$/erg s$^{-1}$) =  2.2 log(SFR/$M_\odot$ yr$^{-1}$) + 39.1.
   We further found a statistically significant super-linear correlation ($\beta = 1.34$ $\pm$ 0.86) between $L^{\rm gas}_{0.5\text{–}2\,\mathrm{keV}}$/SFR and HCN(4–3)/CO(1–0) intensity ratio, whereas no such trend is seen for HCO$^+$(4-3)/CO(1-0) or CO(3-2)/CO(1-0). These findings indicate that AGN feedback does not dominate star formation regulation on sub-kiloparsec scales, and that the excitation of dense gas traced by HCN~(4-3) may be more directly influenced by high-energy feedback processes compared to HCO$^+$ (4-3) and CO (3-2).}
   {}


   \maketitle

\nolinenumbers

\section{Introduction}
The process of star formation within galaxies is a key focus of modern astrophysics. Observational studies on galactic scales, using tracers such as H$\alpha$, H{\sc i}, and CO lines, have revealed a power-law relationship between surface densities of star formation rate (SFR) and total neutral gas, expressed as $\Sigma_{\rm SFR} \varpropto \Sigma_{\rm gas}^{N}$, with an index of $N$ $\approx$ 1.4 \citep{Kennicutt1998ApJ...498..541K,Kennicutt2012ARA&A..50..531K}. This empirical relation is widely known as the Kennicutt–Schmidt (K–S) law. Subsequent studies indicate that stars primarily form from molecular gas \citep[e.g.,][]{Bigiel2008AJ....136.2846B,Liu2015ApJ...805...31L}.

Star formation has been proposed to only occur in molecular regions with column densities higher than the density that exceeds threshold regions in molecular clouds \citep[A$_{\rm v}$ $>$ 8,][]{Lada2010ApJ...724..687L,Heiderman2010ApJ...723.1019H}. 
While the turbulence-regulated model suggests that the formation of stars is governed by the balance between gravitational collapse and the pressure support provided by turbulence within the gas \citep[e.g.,][]{Krumholz2005ApJ...630..250K,Krumholz2007ApJ...669..289K,Federrath2012ApJ...761..156F,Usero2015AJ....150..115U}. 
Observations show that gas tracers with a critical density ($n_{\rm crit}$) $> 10^4$ cm$^{-3}$  (e.g., rotational transitions of HCN, HCO$^{+}$, CS) exhibit a closer correlation with active star-forming regions than the lower transitions of CO 
\citep[e.g.,][]{Wu2005ApJ...635L.173W,Lada2010ApJ...724..687L,Lada2012ApJ...745..190L,Evans2014ApJ...782..114E,Zhang2014ApJ...784L..31Z}. 
In particular, a tight linear correlation between total infrared (TIR) and HCN~(1-0) luminosity has been established based on a large sample of local galaxies \citep{Gao2004ApJS..152...63G,Gao2004ApJ...606..271G}. Moreover, this relationship also holds at high redshifts \citep[e.g.,][]{Gao2007ApJ...660L..93G} and individual Galactic clumps \citep{Wu2005ApJ...635L.173W}.

With advancements in telescopes, high-resolution imaging of molecular gas has enabled a more detailed understanding of the star formation process and its surrounding physical environments on resolved scales \citep[e.g.,][]{Schinnerer2013ApJ...779...42S,Leroy2021ApJS..257...43L,Lee2021AAS...23833104L}.
However, the connection between star formation and its physical environment remains not fully understood. Several studies reported a breakdown of the K–S relation at resolved scales \citep{Onodera2010ApJ...722L.127O,Xu2015ApJ...799...11X,Nguyen-Luong2016ApJ...833...23N}, possibly due to variations in the evolutionary stages of giant molecular clouds  \citep{Onodera2010ApJ...722L.127O}. 
On the one hand, stellar and supernova feedback can significantly influence star formation efficiency by heating gas \citep{Cox2006ApJ...650..791C,Springel2005ApJ...622L...9S,Choi2015MNRAS.449.4105C}. Notably, diffuse 2–10 keV X-ray emission from the hot interstellar medium serves as a crucial diagnostic of stellar feedback, exhibiting a super-linear correlation with the star formation rate in nearby star-forming galaxies  \citep{Ranalli2003A&A...399...39R,Grimm2003MNRAS.339..793G,Gilfanov2004MNRAS.347L..57G,Zhang2024ApJ...967L..25Z,Zhang2025ApJ...988..263Z}.

On the other hand, the role of active galactic nucleus (AGN) feedback in regulating star formation remains controversial. 
While some studies suggest that AGN activity can enhance star formation and boost the SFR (e.g., \citealt{Santini2012A&A...540A.109S,
Mountrichas2024A&A...683A.143M}), others propose that AGN feedback may suppress star formation by heating or expelling cold gas from the host galaxy, resulting in a lower SFR compared to non-AGN systems (e.g., \citealt{Shimizu2015MNRAS.452.1841S}).

Understanding the link between star formation and dense molecular gas with $n > 10^6$ cm$^{-3}$ requires observations of high-$J$ transitions of HCN and HCO$^+$ \citep{Shirley2015PASP..127..299S}. 
To bridge the observational gap between Galactic and extragalactic studies of spatially resolved star formation, the mapping the dense molecular gas in the strongest star-forming galaxies (MALATANG) project was initiated, which is the first systematic survey mapping HCN $J = 4-3$ and HCO$^{+}$ $J = 4-3$ line emissions across 28 nearby IR-bright galaxies using the James Clerk Maxwell Telescope (JCMT). A linear correlation between dense gas luminosity and total infrared luminosity on sub-kpc scales was reported by \citet{Tan2018ApJ...860..165T}, based on observations of six galaxies from the MALATANG survey. 

\object{NGC~1068}, one of the targets in the MALATANG project, is a nearby typical Seyfert 2 galaxy at a distance of $D = 15.7$ Mpc \citep[1$^{\prime \prime}$ = 76 pc,][]{Tan2018ApJ...860..165T}. 
NGC~1068 is known to host an AGN, which is surrounded by a circumnuclear starburst ring \citep{Simkin1980ApJ...237..404S}. 
As one of the nearest and isolated luminous infrared galaxies (LIRGs), it has a high global SFR of $<$ 44 $M_{\odot}$ yr$^{-1}$ \citep{Howell2010ApJ...715..572H}, which represents an upper limit because of AGN contamination, while the disk SFR is 3.2 $M_{\odot}$ yr$^{-1}$ \citep{Nagashima2024ApJ...974..243N}. This makes it an ideal laboratory for investigating how AGN feedback affects dense molecular gas emission and star formation. However, previous studies of dense molecular gas in NGC~1068 (e.g., \citealt{Schinnerer2000ApJ...533..850S,Krips2011ApJ...736...37K,Wang2014ApJ...796...57W,Sanchez-Garcia2022A&A...660A..83S,Gamez2025A&A...699A.187G}) have primarily focused on high-resolution observations of the starburst ring and the central circumnuclear disk (CND) within approximately 1 kpc. Investigations of dense gas in the larger galactic disk and spiral arms remain relatively limited.

This paper is organized as follows. In Sect. \ref{sec:obser}, we describe the observations and data reduction procedures. In Sect. \ref{sec:result}, we present the main results, including the integrated-intensity maps, the correlations between line luminosity and infrared luminosity, as well as between X-ray luminosity and the SFR.
In Section \ref{sec:discussion}, we discuss the influence of AGN activity on infrared, X-ray, and molecular gas emission, and examine the correlations between dust temperature and molecular line ratios, as well as between the X-ray luminosity/SFR ratio and line ratios. Section \ref{sec:summary} summarizes our main conclusions.

\section{Observations and data reduction}
\label{sec:obser}
In this section, we present the observations and data reduction of NGC~1068 in the context of the MALATANG project, as well as the ancillary data used in this work. The fundamental properties of NGC~1068 are summarized in Table \ref{tab:1}.

\begin{table}[h]
 \caption[]{\label{nearbylistaa2}Parameters of NGC 1068.}
 \label{tab:1}
\begin{tabular}{lccc}
 \hline \hline
Parameter &  Value &  Reference \\
 \\ \hline
Coordinate (J2000) &   02h42m40.711s,-00d00m47.81s &  (1) \\
Distance & 15.7 Mpc & (2) \\
Inclination & 41$\pm$2 deg   & (3) \\
$V_{\rm hel}$  & 1127$\pm$3 km s$^{-1}$  & (3) \\
SFR & $<$ 44.29 $M_{\odot}$ yr$^{-1}$  & (4) \\
Type &  Seyfert 2  & (5) \\
\hline
\end{tabular}
\tablebib{(1)~\citet{Capetti1997ApJ...476L..67C};
(2) \citet{Tan2018ApJ...860..165T}, calculated using H$_0 =71$ km s$^{-1}$ Mpc$^{-1}$ corrected for the Virgo infall motion; (3) \citet{Garcia-Burillo2014A&A...567A.125G}; 
(4) \citet{Howell2010ApJ...715..572H}, which should be regarded as an upper limit because of possible AGN contamination; (5) \citet{Osterbrock1993ApJ...414..552O}.
}
\end{table}

\subsection{Observations of HCN~(4-3) and HCO$^{+}$ (4-3)}
\label{subsec:observation} 

The HCN ($4-3$) and HCO$^{+}$ ($4-3$) data of NGC~1068 were obtained from the MALATANG project (program codes: M16AL007 and M20AL022), observed with the Heterodyne Array Receiver Program (HARP) receiver onboard the JCMT telescope. HARP has 16 receptors laid out with 4 $\times$ 4 grids, and two of them (H13, H14) were non-operational during the observation. The receiver backend used is an Auto-Correlation Spectral Imaging System (ACSIS) spectrometer. Observations were carried out in jiggle-chop mode, with each receptor jiggled in a 3 $\times$ 3 patterns and a beam spacing of 10$^{\prime \prime}$. 
The observation was taken in 2015 and 2016, with total integration times of 250 min for HCN ($4-3$), and 287 min for HCO$^{+}$ ($4-3$). The full width at half maximum (FWHM) is approximately 14$^{\prime \prime}$ at 345 GHz. Further observational details can be found in \citet{Tan2018ApJ...860..165T}.

\begin{figure}
    \centering
    \includegraphics[width=0.74\linewidth]{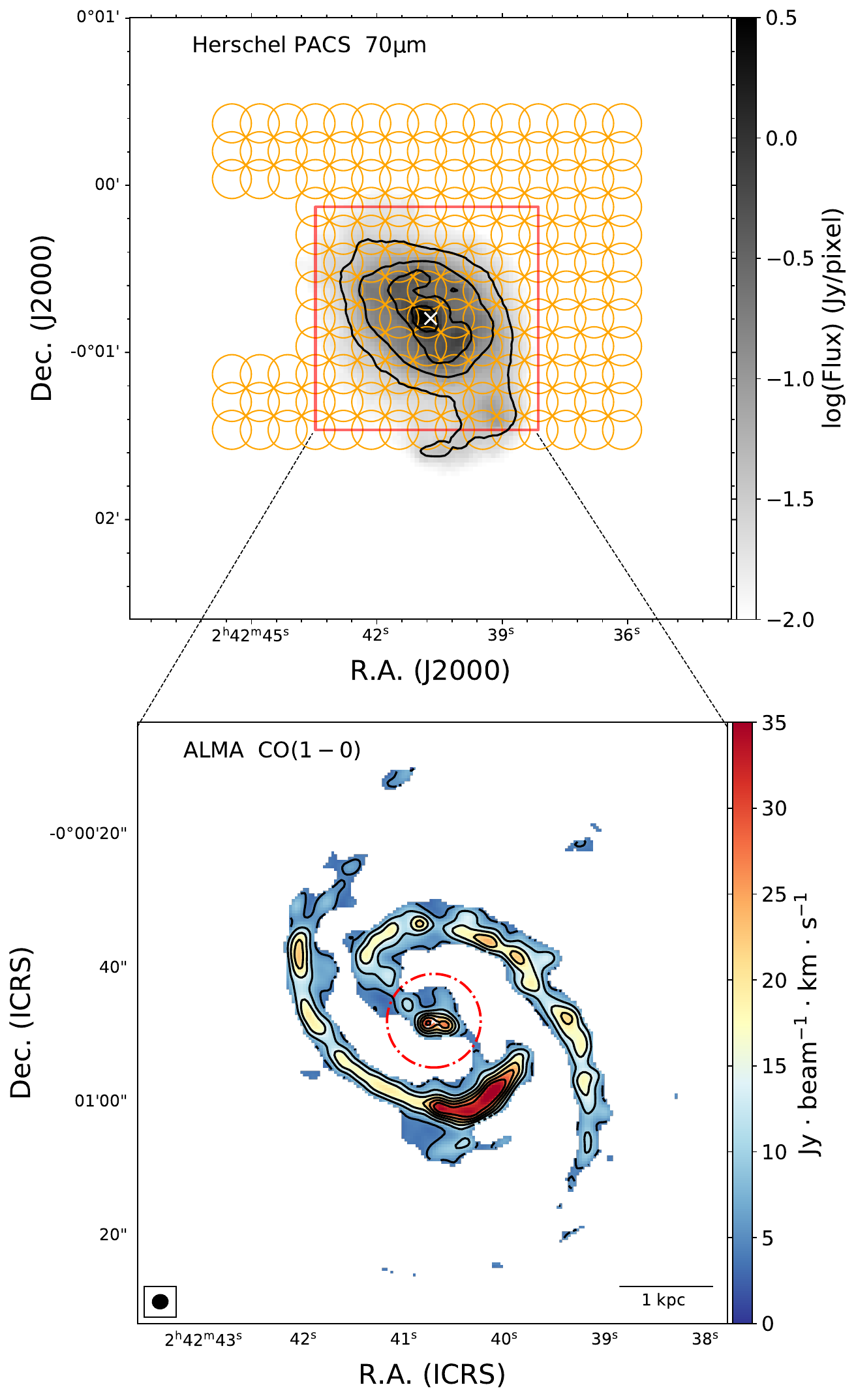}
    \caption{Upper panel: JCMT observing positions of NGC~1068 overlaid on Herschel PACS~70 $\mu$m greyscale map. The black contours indicate the 70~$\mu$m continuum, starting from 100~$\sigma$ and increasing in steps of 3~$\sigma$, where 1~$\sigma = 3\times10^{-4}$~Jy/pixel.
    The white cross denotes the position of the central AGN in the galaxy. Orange circles indicate the observed positions using jiggle mapping mode. The red square shows the area (1.5$^{\prime}$ $\times$ 1.5$^{\prime}$) studied in this work. 
    Bottom panel: A zoom-in view of the central region obtained from the ALMA high-resolution CO~(1–0) map (project ID: 2018.1.01684.S, \citealt{Saito2022ApJ...935..155S}; see also \citealt{Nakajima2023ApJ...955...27N}). Black contours represent CO~(1–0) integrated intensity levels, ranging from 4$\sigma$ to 20$\sigma$ in steps of 4$\sigma$, where 1$\sigma$ = 0.25~Jy~beam$^{-1}$~km~s$^{-1}$. 
    The red dotted-dashed circle here represents the JCMT beam size of $14^{\prime \prime}$ (approximately 1.1~kpc).}
    \label{fig:observation}
\end{figure}

Fig. \ref{fig:observation} shows the observing positions (marked by orange circles) overlaid on a Herschel/PACS 70-$\mu$m image. The MALATANG observations cover the majority of the 70 $\mu$m emission in NGC~1068. Our study area, outlined by a red square, has a map size of 90$^{\prime \prime}$ $\times$ 90$^{\prime \prime}$. Outside this region, HCN~(4-3) and HCO$^{+}$~(4-3) are not detected, with 1$\sigma$ upper limits of $\sim$ 5 mK.
To better illustrate the link between the physical environment and molecular gas distribution, we provide a zoom-in view of the studied area, presented alongside the ALMA high-resolution CO ($1-0$) map in Fig. \ref{fig:observation}. Both the starburst ring (SB ring) and the CND are clearly visible, which are also seen in the arcsec-resolution $^{13}$CO ($1-0$) map \citep{Tosaki2017PASJ...69...18T}.

\subsection{Data reduction}
\label{subsec: Data Reduction}
Data reduction for the HCN $(4-3)$ and HCO$^+$ $(4-3)$ observations was conducted using both the Continuum and Line Analysis Single-dish Software (\texttt{CLASS}) \footnote{\url{http://www.iram.fr/IRAMFR/GILDAS/}}, which is part of the \texttt{GILDAS} package developed by IRAM, and \texttt{STARLINK}, a package developed for JCMT data reduction. 

We first identified and flagged bad sub-scans through visual inspection using GAIA, a visualization tool within the \texttt{STARLINK} software suite. Flagged sub-scans were excluded by applying masks via the \texttt{chpix} task. In addition, high-noise regions located at the spectral edges of each sub-scan were removed to improve data quality.
Following this, the data were converted from the \texttt{STARLINK} N-Dimensional Data Format (NDF) format to the \texttt{GILDAS/CLASS} format. Reduction procedures were then applied individually to each sub-scan using the \texttt{CLASS} software. 
Spikes at the edges of the spectra and corrupted channels were replaced with Gaussian noise to minimize artifacts.
Baseline subtraction was performed using either first- or second-order polynomial fitting, depending on the quality of the spectra. For positions with a signal-to-noise ratio (S/N) less than 3, the CO spectra were used to define appropriate baseline windows. All spectra were smoothed to a velocity resolution of 26~km~s$^{-1}$.
The processed data were then converted back to the NDF format for further reduction using the \texttt{ORAC-DR} pipeline, an automated data reduction procedure developed specifically for JCMT observations within the \texttt{STARLINK} software suite  \citep{Jenness2015MNRAS.453...73J}. 

\subsection{Ancillary data}
\label{subsec:Ancillary Data}
To investigate the extended gas properties, we use ancillary CO(3–2) and CO~(1–0) data, while archival infrared and X-ray observations are used to derive the infrared and X-ray luminosities.

\subsubsection{CO~(3-2) and CO~(1-0) data}
The CO $(3-2)$ data were retrieved from the JCMT archive (project ID: M09BC05) and were observed using HARP and ACSIS, with a FWHM beam size of approximately 14$^{\prime \prime}$, comparable to the MALATANG data.
This observation used scan mode, also known as raster mode, which is the most efficient observing mode for large maps. During the scans, the HARP array was rotated at a specific angle, resulting in a sampling spacing of 10.3$^{\prime \prime}$. 

The CO $(1-0)$ data were obtained from the Nobeyama CO Atlas of Nearby Spiral Galaxies\footnote{\url{https://www.nro.nao.ac.jp/~nro45mrt/html/COatlas/}}, using the NRO 45-m Telescope \citep{Kuno2007PASJ...59..117K}. The beam size FWHM is about 15$^{\prime \prime}$, which is comparable to our MALATANG data. 

The reduction of the CO ($1\text{–}0$) data was performed using the CLASS software, while the CO ($3\text{–}2$) were reduced following the standard JCMT data reduction procedures. The CO ($1\text{–}0$) and CO ($3\text{–}2$) maps were regridded to a common pixel size of 10$^{\prime\prime}$, consistent with the MALATANG data.

\subsubsection{Infrared data}
\label{subsubsec:Infrared data}
To investigate the infrared characteristics of NGC~1068, we include the infrared photometric data at 70~$\mu$m, 160~$\mu$m, and 850~$\mu$m bands.
The 70 $\mu$m and 160 $\mu$m data were obtained from Herschel/PACS, with FWHM angular resolutions of approximately 5.8$^{\prime \prime}$ at 70 $\mu$m and 12$^{\prime \prime}$ at 160 $\mu$m, respectively.
To match the resolution of the MALATANG observations, the \emph{Herschel} maps were convolved using the convolution kernels provided by \citet{Aniano2011PASP..123.1218A} to achieve a common beam size of 14$^{\prime\prime}$.
The maps were then scaled by a factor of $1.133 \times (14/\text{pixel size})^2$ to convert the flux units from Jy to Jy beam$^{-1}$, where the pixel size is in arcseconds.

The 850 $\mu$m data were obtained from the Submillimetre Common-User Bolometer Array 2 (SCUBA2) on the JCMT, with a FWHM of approximately 14.6$^{\prime \prime}$. 
To convert the flux units from pW to Jy, we used the standard Flux Conversion Factor of 2.13 $\pm$ 0.12 Jy pW$^{-1}$ arcsec$^{-2}$ for the 850 $\mu$m data.

\subsubsection{X-ray data}
We used the Chandra observation conducted in 2000 (ObsIds 344) with an exposure time of 47.44 ks and a beam size of approximately 0.5$^{\prime \prime}$. The data were reprocessed using CIAO v4.17 and CALDB v4.11.6.
To obtain the emission of diffuse X-ray-emitting gas, we identified point sources by the wavelet-based source detection algorithm \texttt{wavdetect} \citep{Freeman2002ApJS..138..185F} in the energy range of 0.5-7 keV and on the wavelet scales of 1, $\sqrt{2}$, 2, 2$\sqrt{2}$, 4, 4$\sqrt{2}$, and 8. The 90\% energy-encircled ellipse was adopted for the source region based on the point spread function (PSF) shape. To minimize ring-like artifacts from the PSF wings, we manually increased the ellipse sizes for certain bright sources.
The point source regions were then subtracted from the X-ray map. Given the Seyfert 2 AGN in the galaxy, we also removed the central region (approximately 8$^{\prime \prime}$ radius) of NGC~1068. The spectra of hot gas were extracted from regions corresponding to the size of the JCMT beam (14$^{\prime \prime}$) at each observation position, using the CIAO routine \texttt{specextract}, with the background region defined on the same CCD chip, excluding point sources and the target regions. 
We fitted the hot gas spectra using the XSPEC v.12.11.0 \citep{Arnaud1996ASPC..101...17A} software with an absorbed Astrophysical Plasma Emission Code (APEC; \citealt{Smith2001ApJ...556L..91S}). The fitted X-ray luminosities were corrected for both Galactic and intrinsic absorption.

\section{Result}
\label{sec:result}
\subsection{Integrated intensity images}
\label{subsec:Integrated intensities images}
\begin{figure*}
    \centering
    \includegraphics[width=\linewidth]{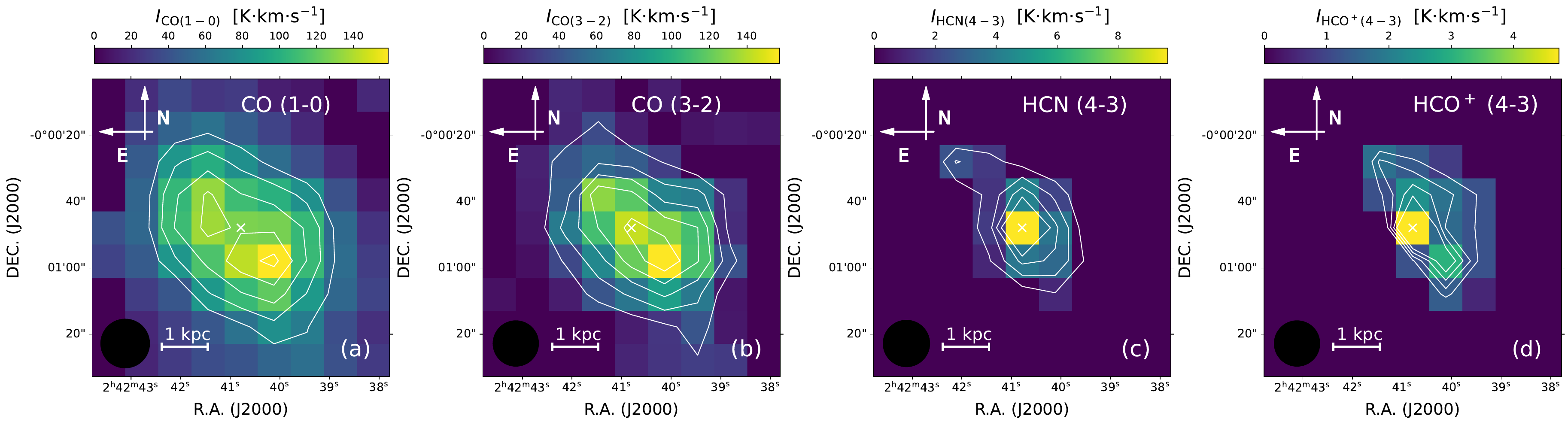}
    \caption{Integrated intensity (moment 0) maps of molecular gas tracers, shown with a pixel size of 10$^{\prime \prime}$. (a): CO ($1-0$) map with contours at levels of 15 to 35 $\sigma$ in steps of 5 $\sigma$ ($\sigma = 4.34$ K~km~s$^{-1}$); (b): CO ($3-2$) map with contours from 13 to 57 $\sigma$ in steps of 11 $\sigma$ ($\sigma = 2.39$ K~km~s$^{-1}$); (c): HCN ($4-3$) map with contours from 3 to 12 $\sigma$ in steps of 3 $\sigma$ ($\sigma = 0.38$ K km s$^{-1}$); (d): HCO$^{+}$ ($4-3$) map with contours from 3 to 6 $\sigma$ in steps of 1 $\sigma$ ($\sigma = 0.41$ K km s$^{-1}$). The white cross marks the location of the AGN in NGC~1068. Pixels with S/N below 3 are masked. The beam size and scale bar are shown in the lower-left corner of each panel.}
    \label{fig:Integral_intensity_diagram}
\end{figure*}

Figure \ref{fig:Integral_intensity_diagram} displays the spatial distributions of the integrated intensities for HCN ($4-3$), HCO$^+$ ($4-3$), CO ($3-2$), and CO ($1-0$). 
The position of the central AGN is marked with a white cross. 
All four tracers exhibit elongated morphologies along the southwest–northeast direction, aligning with the major axis of NGC~1068. 
The molecular gas traced by HCN~($4\text{–}3$) and HCO$^+$~($4\text{–}3$) is significantly more compact than the distributions seen in CO~($1\text{–}0$) and CO~($3\text{–}2$). 
We note that the distribution of HCO$^+$ ($4-3$) is highly asymmetric compared to the other tracers, with more prominent emission in the western region than the eastern region.
Both HCN ($4-3$) and HCO$^+$ ($4-3$) emissions peak at the center, coinciding with the position of the AGN.

In contrast, CO~($1\text{–}0$) and CO~($3\text{–}2$) emissions exhibit their peak intensities in the southwestern region.
Using interferometric data, \citet{Tsai2012ApJ...746..129T} reported that CO~($3\text{–}2$) emission in NGC~1068 typically peaks at the nucleus, whereas CO~($1\text{–}0$) emission is predominantly distributed along the spiral arms. However, at our current resolution, the spatial distribution of CO~($3\text{–}2$) is similar to that of CO~($1\text{–}0$).  In particular, the southwestern region shows the strongest CO~($3\text{–}2$) and CO~($1\text{–}0$) emissions, consistent with the SB knot within the pseudo-ring structure revealed in the ALMA CO~($1\text{–}0$) map (see Fig. \ref{fig:observation}). Spectral maps of these lines (Figure A1 and A2) are provided in Zenodo \href{url}{https://doi.org/10.5281/zenodo.17741281}.

\subsection{Line and infrared luminosities}
\label{subsec:Luminosity}
We calculate the line luminosities using the method described by \citet{Solomon1997ApJ...478..144S}:
\begin{equation}
    L_{\rm dense}^{'} = 3.25 \times 10^7 (S\bigtriangleup_\nu)(\nu_{\rm obs})^{-2} \times (D_{\rm L})^{2}(1+z)^{-3}     \ ,
\label{equ:2}
\end{equation}
where $S\bigtriangleup_v$ is the velocity-integrated flux density in units of Jy km s$^{-1}$, and $v_{\rm obs}$ is the observed line frequency in units of GHz. $D_{\rm L}$ is the luminosity distance (see Table \ref{tab:1}). To convert line intensity to flux density, we used the conversion factor \(S/T_{\rm mb} = 15.6/\eta_{\rm mb} \), where $\eta_{\rm mb}$ is the main beam efficiency.
We adopt $\eta_{\rm mb}(\rm JCMT)$ = 0.64 and $\eta_{\rm mb}(\rm NRO~45m)$ = 0.40.

The uncertainties of line intensity were derived from 
\begin{equation}
    \sigma_I = T_{\rm rms} \sqrt{\bigtriangleup v_{\rm line} \bigtriangleup v_{\rm res}} \sqrt{1+\bigtriangleup v_{\rm line} / \bigtriangleup v_{\rm base}} \ ,
    \label{equ:3}
\end{equation}
where $T_{\rm rms}$ is the RMS main-beam temperature, $\bigtriangleup v_{\rm res}$ is the spectral velocity resolution, $\bigtriangleup v_{\rm line}$ is the velocity range of the emission line,  
and $\bigtriangleup v_{\rm base}$ is the velocity range used to fit the baseline \citep{Gao1996,Matthews2001ApJ...549L.191M}. 
We adopt the velocity ranges of CO~($3\text{–}2$) as a reference for determining the spectral ranges used in the baseline fitting of the HCN~($4\text{–}3$) and HCO$^+$~($4\text{–}3$) lines at outer regions, since CO~($3\text{–}2$) emission is more spatially extended and exhibits similar kinematic properties as HCN~($4\text{–}3$) and HCO$^+$~($4\text{–}3$).
Note that an additional 10\% uncertainty is included to account for systematic errors in the absolute flux calibration.

In order to calculate the total infrared luminosities (from 8 $\mu$m to 1000 $\mu$m) of NGC~1068, we estimate the resolved and integrated $L_{\rm TIR}$ from \emph{Herschel} bands with combined tracers \citep{Galametz2013MNRAS.431.1956G}, using the equation below:
\begin{equation}
    L_{\rm TIR} = \sum c_i \nu L_\nu(i) L_{\odot} \ ,
\label{equ:4}
\end{equation}
where \(\nu L_\nu(i) \) is the resolved luminosity in units of \(L_{\odot} \), and \(c_i \) is the calibration coefficients for the combination of different bands. For NGC~1068, the equation is \(L_{\rm TIR} = (1.010 \pm 0.023) \ \nu \emph{L}_\nu(70\mu m)+ (1.218 \pm 0.017) \ \nu \emph{L}_\nu(160\mu m)\).  
The uncertainty in the $L_{\rm TIR}$ calibration derived from the combination of multiple infrared bands is adopted to be 25\% \citep{Galametz2013MNRAS.431.1956G}. Additionally, we assume a 5\% flux calibration uncertainty, following \citet{Balog2014ExA....37..129B}.

The intensities, line luminosities of the four molecular tracers, and the total infrared luminosities for individual resolved regions are listed in Appendix \ref{Appendix: C}.  
All uncertainties are estimated from the measurements using equation \ref{equ:3}. For positions with signal-to-noise ratios below 3, only 3$\sigma$ upper limits are provided in the table.

\subsection{Relation between dense gas tracers and IR luminosities}
\label{Subsec:gas_IR_luminosity}
The dense gas usually refers to molecular gas with $n_{\rm H_2}$ $>$ 10$^4$ cm$^{-3}$ \citep{Lada1992ApJ...393L..25L}. 
HCN~(4–3) and HCO$^+$~(4–3) have critical densities approximately 2.3 $\times$ 10$^7$ cm$^{-3}$ and 3.2 $\times$ 10$^6$ cm$^{-3}$ at 20~K under optically thin assumptions, respectively \citep{Shirley2015PASP..127..299S}. 
Given their high critical densities, we consider HCN~(4–3) and HCO$^+$~(4–3) as tracers of dense molecular gas in this study.

\begin{figure*}
    \centering
    \includegraphics[width=\linewidth]{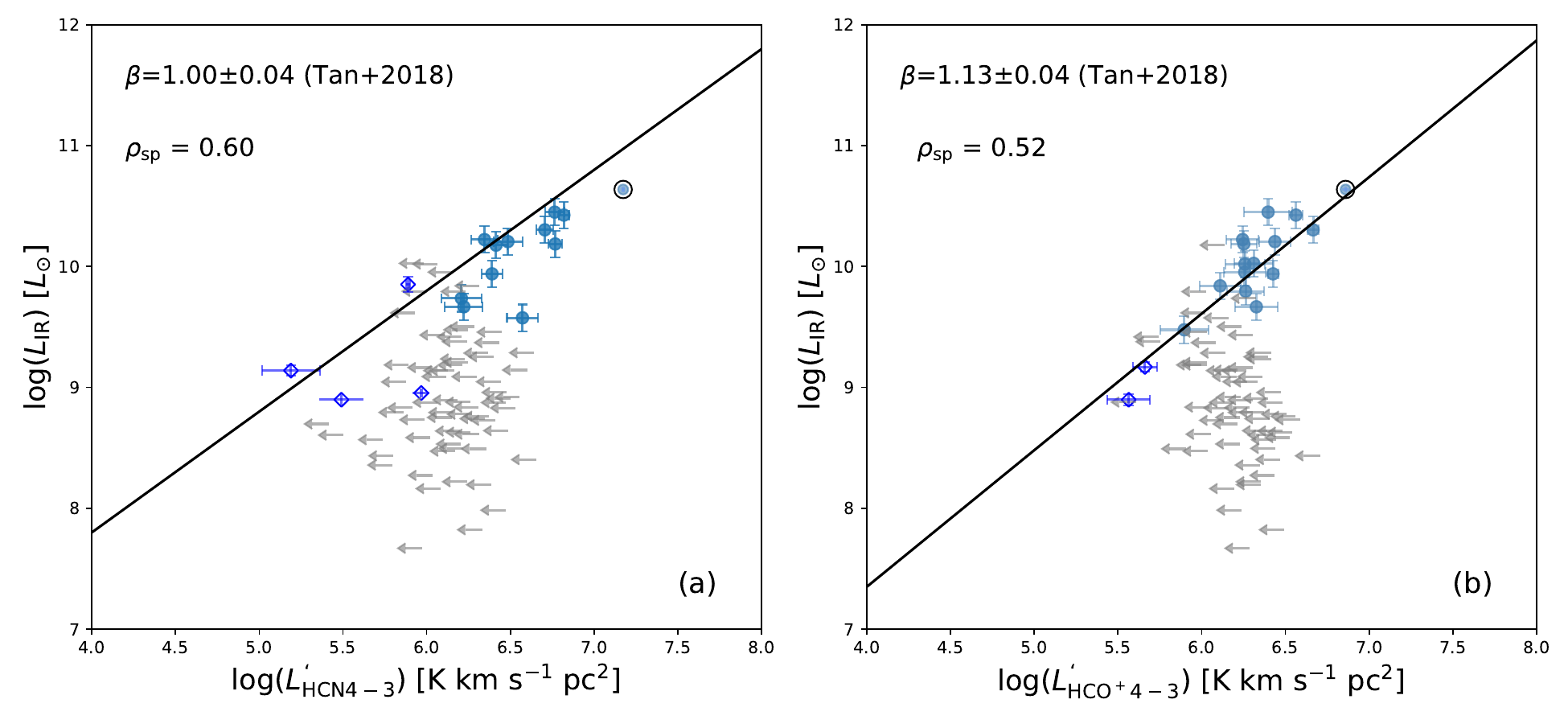}
    \caption{Relations between IR luminosities and dense molecular line luminosities in logarithmic scale. (a) $L_{\rm IR} - L^{'}_{\rm HCN (4-3)}$; (b) $ L_{\rm IR} - L^{'}_{\rm HCO^+ (4-3)}$.
    Detected data are shown as solid points, data points obtained from stacked spectra are shown as diamonds, and upper limits are marked by grey leftward arrows. 
    The black circles highlight the central region of NGC~1068, which hosts the AGN.
    The Spearman rank correlation coefficients ($\rho_{\rm sp}$) are displayed in each panel.
    The solid lines and their slopes ($\beta$) represent the the best-fit results from Bayesian regression, as adopted from \citet{Tan2018ApJ...860..165T}.
    }
    \label{fig:HCN_HCO_IR}
\end{figure*}

We examine the correlations between the luminosities of molecular gas and the IR luminosities in Fig. \ref{fig:HCN_HCO_IR}.
To improve the S/N ratio in the outer regions, we average the spectra from pixels at the same radial distance to obtain stacked data, which are marked by diamonds in Fig. \ref{fig:HCN_HCO_IR}. The HCN ($4-3$) and HCO$^{+}$ ($4-3$) emissions are detected at approximately 3 $-$ 8$\sigma$ levels in the outer disk region (approximately 2.2 $-$ 3.2 kpc radial distance) by stacking spectra from different radial positions (see Figure A3 via Zenodo \href{url}{https://doi.org/10.5281/zenodo.17741281}). 
The $L_{\rm IR}-L'_{\rm dense}$ correlation is consistent with the fitting results reported by \citet{Tan2018ApJ...860..165T}, as indicated by the black solid line in Fig. \ref{fig:HCN_HCO_IR}.
Notably, despite the influence of AGN activity in the central regions of NGC~1068, all data points still follow the 
linear $L_{\rm IR} - L^{\prime}_{\rm dense}$ correlation. The impact of the AGN is discussed in Section \ref{subsec:AGN affect}.

\subsection{Relation between $L^{\rm gas}_{\rm 0.5 - 2\,keV}$ and SFR}
\label{Subsec:Lx-SFR}
\begin{figure}
    \centering
    \includegraphics[width=\linewidth]{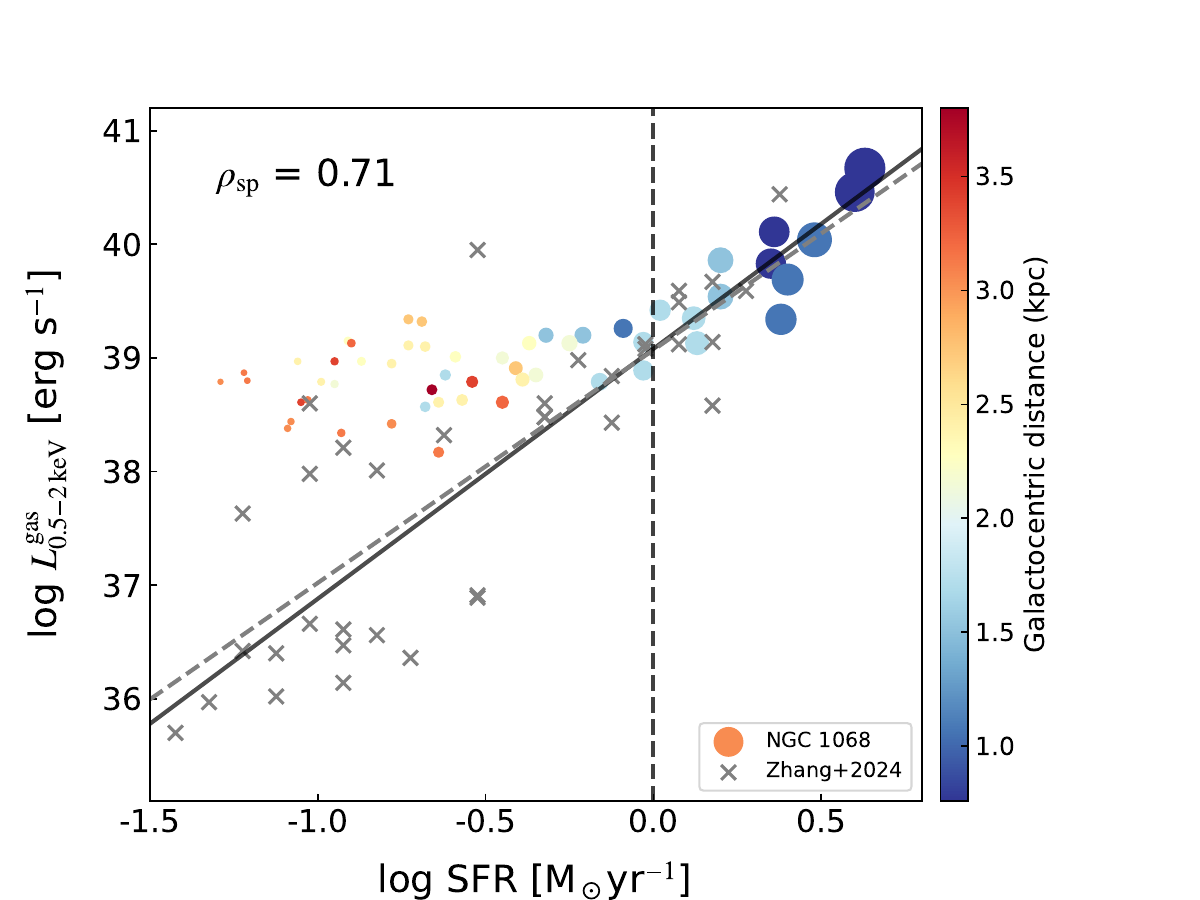}
    \caption{Relation between SFR and thermal X-ray luminosity of hot gas on sub-kpc scale. Color scale represents the distance from the galactic center, and the marker size represents the SFR surface density. The central data point containing the AGN has been exclude here. The vertical black dashed line marks SFR = 1~$M_{\odot}$~yr$^{-1}$, corresponding to $\Sigma_{\rm SFR}$ = 8.2 $\times$ 10$^{-6}$ $M_\odot$ yr$^{-1}$ kpc$^{-2}$. The black solid line represents the best-fit scaling relation as expressed in Eq. \ref{equ:Lx_SFR}, derived from the central 12 data points with SFR $>$ 1~$M_{\odot}$~yr$^{-1}$. The Spearman rank correlation coefficient ($\rho_{\rm sp}$) is given in the panel. The gray crosses represent data from the central regions of nearby galaxies \citep{Zhang2024ApJ...967L..25Z}, while the gray dashed line indicates the corresponding best-fit relation for these galaxies.}
    \label{fig:Lx-SFR}
\end{figure}

The diffuse soft X-ray emission (0.5–2 keV band) from hot gas produced by stellar feedback is found to correlate with the SFR, particularly in starburst regions \citep[e.g.,][]{Ranalli2003A&A...399...39R,Grimm2003MNRAS.339..793G,Zhang2024ApJ...967L..25Z}. We explore the relation between the diffuse thermal X-ray luminosity ($L^{\rm gas}_{\rm 0.5 - 2\,keV}$) of the hot gas and the SFR (Fig. \ref{fig:Lx-SFR}).
Assuming that the infrared luminosity is dominated by star formation, we estimate the SFR from the total infrared luminosity using the relation given by \citet{Kennicutt1998ApJ...498..541K}:

\begin{equation}
    (\rm {SFR}/ M_{\odot} \rm{yr}^{-1}) = 1.50 \times 10^{-10} (\emph{L}_{\rm TIR}/ \emph{L}_{\odot}) \ ,
\end{equation}
where the $L_{\rm TIR}$ is the total infrared luminosity estimated by equation (\ref{equ:4}). Note that this relation is typically used to estimate the global SFR of entire galaxies. However, in this study, we apply it to spatially resolved regions under the assumption that the relation holds at sub-kpc scales. The SFR surface density ($\Sigma_{\text{SFR}}$) is calculated as the SFR within a region divided by its physical area:
$\Sigma_{\text{SFR}} = \frac{\text{SFR}}{A}$,
where $A$ is the area over which the SFR is measured, in units of $\text{kpc}^2$.

A bimodal pattern is found in this correlation, distinguishing the central region, where SFR $>$ 1 $M_{\odot}$ yr$^{-1}$ (i.e., $\Sigma_{\rm SFR}$ $>$ 8.2 $\times$ 10$^{-6}$ $M_\odot$ yr$^{-1}$ kpc$^{-2}$) from the outer disk, where SFR $<$ 1 $M_{\odot}$ yr$^{-1}$ (i.e., $\Sigma_{\rm SFR}$ $<$ 8.2 $\times$ 10$^{-6}$ $M_\odot$ yr$^{-1}$ kpc$^{-2}$), similar to the findings reported in M51 \citep{Zhang2025ApJ...978L..15Z}. 
Variations between the central and disk regions indicate spatially distinct gas heating mechanisms within the galaxy. Central hot gas may primarily be heated by the combined influence of AGN activity and starburst processes.

We used the Python package \emph{Linmix} to perform Bayesian linear regression using Markov Chain Monte Carlo (MCMC) sampling \citep{Kelly2007ApJ...665.1489K}. 
We used only the central 12 data points (with radial distances $<$ 1.7 kpc), where each region exhibits a high star formation rate surface density ($\Sigma_{\rm SFR}$ $>$ 8.2 $\times$ 10$^{-6}$ $M_\odot$ yr$^{-1}$ kpc$^{-2}$) and shows detections of both HCN(4–3) and HCO$^{+}$(4–3).
Using these data, the linear regression yields a super linear relation with a Spearman rank correlation coefficient of 0.71 and
a p-value of 0.01 (black solid line in Fig. \ref{fig:Lx-SFR}):

\begin{equation}
    \log \frac{L^{\rm gas}_{0.5-2 \rm keV}}{\rm erg~s^{-1}} = 2.20(\pm0.73) \log \frac{\rm SFR}{M_{\odot} ~ \rm yr^{-1}} + 39.08(\pm0.28).
    \label{equ:Lx_SFR}
\end{equation}

The X-ray luminosity ranges from 10$^{38}$ to 10$^{41}$ erg~s$^{-1}$, which is broadly consistent with the soft-band luminosities typically found in nearby star-forming galaxies at the sub-kpc scale \citep[e.g.,][]{Zhang2024ApJ...967L..25Z}.
The $L^{\rm gas}_{\rm 0.5 - 2\,keV}$ - SFR relation shows a significant excess over the linear trend, consistent with the findings of \citet{Zhang2024ApJ...967L..25Z}, as indicated by the gray dashed line in Fig. \ref{fig:Lx-SFR}.
This super-linear relation is likely driven by the extreme conditions in galaxy centers, including AGN activity and intense star formation, both of which contribute to the accumulation of hot gas at the galactic center. Furthermore, AGN may influence X-ray and infrared emissions with different efficiencies, which could further contribute to the super-linear correlation.

\section{Analysis and discussion} 
\label{sec:discussion}

\subsection{AGN feedback effects}
\label{subsec:AGN affect}
Both star formation and AGN activity can heat the surrounding dust and gas, leading to enhanced infrared luminosity and elevated molecular line emission (e.g., \citealt{Krips2008ApJ...677..262K,Izumi2013PASJ...65..100I}). In particular, the CND in NGC~1068 is strongly influenced by ionized winds driven by the AGN \citep[e,g.,][]{Garcia-Burillo2014A&A...567A.125G}, which contribute to high temperatures that further enhance the abundances of molecules \citep[e,g.,][]{Harada2010ApJ...721.1570H}. The spatial resolution of our JCMT observations is insufficient to disentangle the AGN, CND, and SB ring. Therefore,
both the central beam (covering the AGN and the CND) and nearby beam-covered regions in our observations are likely influenced by the combined effects of the CND and the AGN. 
As mentioned in Section \ref{subsec:Integrated intensities images}, the emissions of HCN~(4-3) and HCO$^+$~(4-3) are likely enhanced by AGN-driven radiation in the central region of NGC~1068. 
However, AGN activity does not appear to significantly affect the $L_{\rm IR} - L^{\prime}_{\rm dense}$ relation, which remains consistent with the general trend of the star formation law, as presented in Section \ref{Subsec:gas_IR_luminosity}.

A plausible explanation is that the influence of AGN activity in NGC~1068 is localized rather than galaxy-wide.  
\cite{Esquej2014ApJ...780...86E} indicates that AGN radiation has no significant impact on star formation at scales of approximately 100~pc from the nucleus. 
Similarly, a research towards nearby AGN-hosting galaxies by \citet{Esposito2022MNRAS.512..686E} found no significant evidence of AGN influence on the molecular gas properties at kpc scales. 
Although the mechanical feedback from AGN, such as jets and outflows, can extend over several kpc \citep{Garcia-Burillo2014A&A...567A.125G,Saito2022ApJ...927L..32S}, star formation is the dominant contributor to the total infrared luminosity in AGN-embedded systems \citep{Nardini2008MNRAS.385L.130N,Sanders2012ApJS..203....9U}. 
A study of NGC~1068 by \cite{Tsai2012ApJ...746..129T} indicates that while the molecular gas in the nuclear region is influenced by AGN radiation, the gas in the spiral arms is mainly affected by star formation.
As suggested by \citet{Juneau2009ApJ...707.1217J}, HCN can serve as a reliable tracer of the dense gas responsible for star formation, even in the presence of AGNs.
Moreover, the extremely short timescale of the AGN outflow (approximately 0.2 Myr, \citealt{Das2006AJ....132..620D}) suggests that AGN feedback may have only a minor influence at the current stage. 
Alternatively, AGN could affect both the infrared and molecular line emissions in a comparable way, which could explain why the central region still follows the linear $L_{\rm IR} - L^{\prime}_{\rm dense}$ relation.

The AGN is estimated to contribute approximately 10\% of the diffuse X-ray luminosity within the central region of M51 \citep{Terashima2001ApJ...560..139T}. 
\citet{Smith2018AJ....155...81S} demonstrates that the $L^{\rm gas}_{\rm\,0.5\text{–}2\,keV}$/SFR ratio remains unaffected by AGN activity, supporting the interpretation that diffuse soft X-ray emission is primarily powered by feedback from stars and supernovae.
Therefore, the AGN contribution to the soft-band X-ray luminosity is also relatively small in the outer region.
Note that the observed diffuse X-ray emission may also include contributions from faint, unresolved point sources such as X-ray binaries, cataclysmic variables, active binaries, supernova remnants, and young stellar objects. However, these compact sources contribute only minimally to the total soft X-ray emission \citep{Kuntz2010ApJS..188...46K,Wang2021MNRAS.508.6155W,Zhang2025ApJ...988..263Z}.

\subsection{Dust temperature dependence of molecular line ratios}
\label{subsec:dust temperature}

To derive the dust temperatures and column densities in NGC~1068, we perform spectral energy distribution (SED) fitting using infrared data at 70~$\mu$m, 160~$\mu$m, and 850~$\mu$m.
We described the dust radiation by a single modified black-body (MBB) model, thus the observed flux density ($S_\nu$) is related to the Planck function $B_{\nu}(T)$ by the following equation:
\begin{equation}
\centering
     S_{\nu} = B_{\nu}(T) (1 - e^{-\tau_\nu})\Omega \ ,
\end{equation}
where $\tau_\nu$ is the optical depth at frequency $\nu$, and $\Omega$ is the solid angle of the source in steradians.

The total column density $N_{\rm H_2}$ is then given by the equation:
\begin{equation}
    N_{\rm H_2} = \frac{\tau_\nu M_g}{\kappa_\nu \mu m_{\rm H} M_d} \ ,
\end{equation}
where $\mu$ is the mean molecular mass per Hydrogen atom, $m_{\rm H}$ is the Hydrogen mass, and a gas-to-dust ratio of 100 is assumed.
The dust emissivity $\kappa_\nu$ at a given frequency is calculated using the following equation: 
\begin{equation}
    \kappa_\nu = \kappa_{230} (\nu / 230~ \rm GHz)^\beta \ ,
\end{equation}
where $\beta$ is the emissivity index, and $\kappa_{230}$ is the dust opacity at 230~GHz. In this work, we adopt $\kappa_{230}$ = 0.899 cm$^2$g$^{-1}$, following  \citet{Ossenkopf1994A&A...291..943O}.

To ensure consistency in spatial resolution, all infrared maps were convolved to match the beam size (14$^{\prime \prime}$) of the MALATANG data. The SED fitting was then performed on a grid of 9 $\times$ 9 pixels to derive spatially resolved maps of dust temperature and column density. The total flux uncertainty accounts for contributions from photometric measurement errors, a 5\% flux calibration uncertainty, and an additional 5\% systematic uncertainty, following \cite{Balog2014ExA....37..129B}.
Note that background and foreground emissions were not subtracted from the IR images.

In the SED fitting, the emissivity index $\beta$ was set as a free parameter. We obtained an  average value of $\beta = 1.94$ for NGC~1068, which is broadly consistent with the typical range  in the galaxies of 1–2.5 \citep{Chapin2011MNRAS.411..505C} and is close to the average value of approximately 1.78 found in the Galactic disk \citep{Planck2011A&A...536A..25P}.
The SED fitting results at representative positions are presented in Appendix\ref{Appendix: D}.
In the central region, the SED fitting reveals a dust temperature of 32.8$\pm$0.6~K and a column density $N_{\rm H_2}$ of 1.6 $\times$ 10$^{22}$ cm$^{-2}$, which is consistent with the cold dust SED results of NGC~1068 in \cite{Zhou2022ApJ...936...58Z}.

\begin{figure*}
    \centering
    \includegraphics[width=\linewidth]{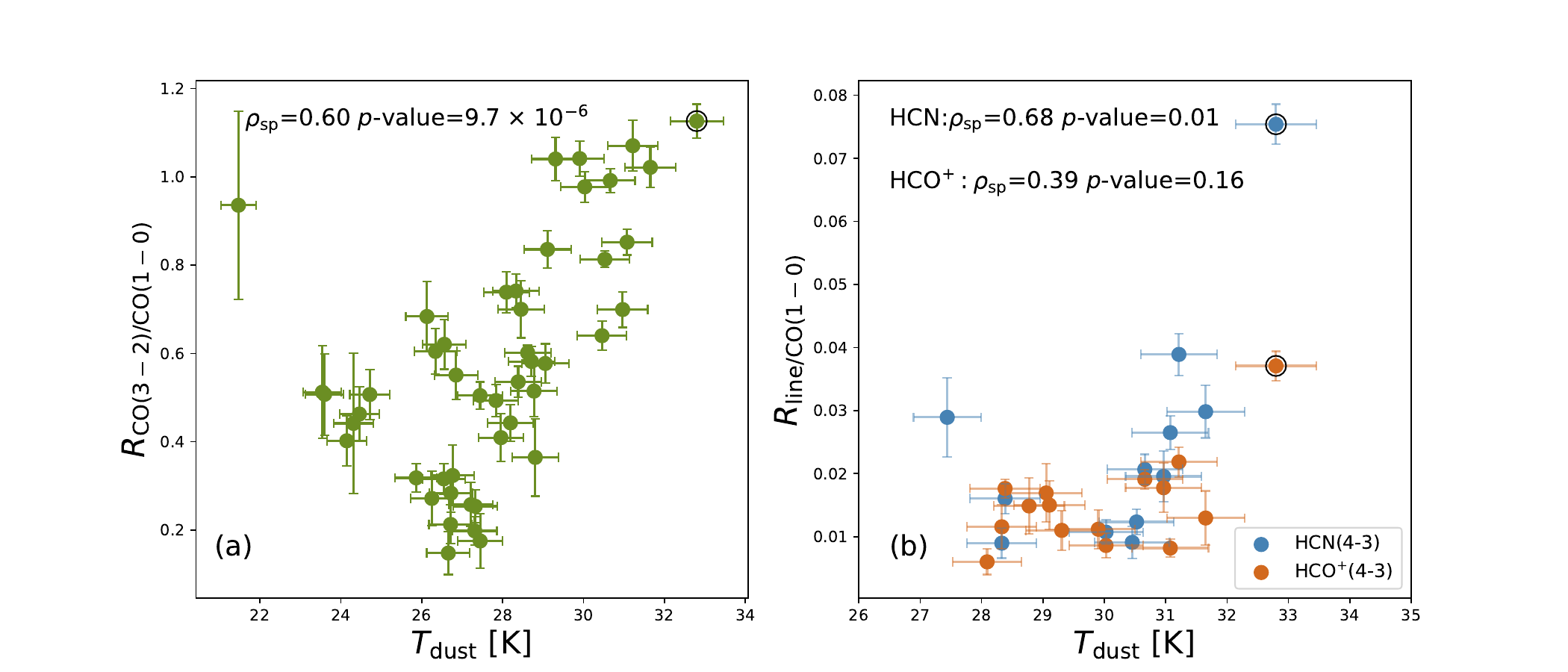}
    \caption{(a): Correlation between dust temperature and gas excitation (traced by $R_{\rm 31}$); (b): Correlation between dust temperature and dense and warm gas fraction, denoted by HCN $(4-3)$/CO $(1-0)$ and HCO$^{+}$ $(4-3)$/CO $(1-0)$ integrated intensity ratios. Black circles denote the central position of NGC~1068, which hosts the AGN. The Spearman rank correlation coefficients are shown at each panel.}
    \label{fig:T_dust_2}
\end{figure*}

We explore the relationships between the molecular line ratios and the dust temperature of NGC~1068, as shown in Fig. \ref{fig:T_dust_2}. The integrated intensity ratios HCN (4–3)/CO (1–0) and HCO$^+$ (4–3)/CO (1–0) 
serve as proxies for dense and warm gas fraction, since HCN~(4–3) and HCO$^{+}$~(4–3) have higher critical densities and excitation requirements compared to CO~(1–0).
Meanwhile, $R_{31}$, defined as $I_{\rm CO(3-2)}/I_{\rm CO(1-0)}$, is a key diagnostic of molecular gas excitation conditions \citep{Mauersberger1999A&A...341..256M,Wang2025A&A...700A..72W}.

Our result reveals a positive correlation between $T_{\rm dust}$ and $R_{\rm 31}$, with a Spearman rank coefficient of $\rho_{\rm sp} = 0.60$ and a $p$-value of $9.7 \times 10^{-6}$.
The mean $R_{\rm 31}$ value of 0.5 $\pm$ 0.1 is consistent with previously reported values for this galaxy \citep{Mauersberger1999A&A...341..256M,Mao2010ApJ...724.1336M}.
Although the central region of NGC~1068 is likely dominated by X-ray-dominated regions (XDRs) \citep{Hailey-Dunsheath2012ApJ...755...57H}, where intense X-ray radiation from the AGN can heat the gas via photoionization and potentially cause dust–gas decoupling, our results show that this region still follows the positive correlation.
This correlation indicates that dust temperature is linked to molecular gas excitation, suggesting that the dust and molecular gas are likely thermally coupled \citep[e.g.,][]{Young1986ApJ...304..443Y,Goldsmith2001ApJ...557..736G}.

A similar positive correlation is found between  the HCN $(4-3)$/CO$(1-0)$ integrated intensity ratio and $T_{\rm dust}$, with a Spearman rank coefficient of $\rho_{\rm sp} = 0.68$ and a $p$-value of 0.01. 
The positive correlation between HCN~(4–3)/CO~(1–0) and $T_{\rm dust}$, particularly in the central region of the galaxy, can be attributed to AGN feedback, which enhances star formation activity and simultaneously heats both the dust and the dense molecular gas traced by HCN~(4–3).
However, there is no clear correlation found between HCO$^+$(4-3)/CO(1-0) intensity ratio and $T_{\rm dust}$. 
HCO$^+$~(4-3) traces denser, more shielded gas components compared to CO~(3-2), whose excitation conditions are less affected by variations in dust temperature (e.g., \citealt{Papadopoulos2014ApJ...788..153P}). Moreover, the fractional abundance of HCO$^+$ is less sensitive to AGN-driven chemistry than that of HCN. These factors  could lead to the weak or absent correlation between HCO$^{+}$(4–3)/CO(1–0) and $T_{\rm dust}$, whereas stronger correlations are found for both $R_{31}$–$T_{\rm dust}$ and HCN(4–3)/CO(1–0)–$T_{\rm dust}$.

\subsection{Relation between $L^{\rm gas}_{\rm\,0.5\text{–}2\,keV}$/SFR and dense gas}
\label{Subsec:Lx_SFR-Line}
\begin{figure*}
    \centering
    \includegraphics[width=\linewidth]{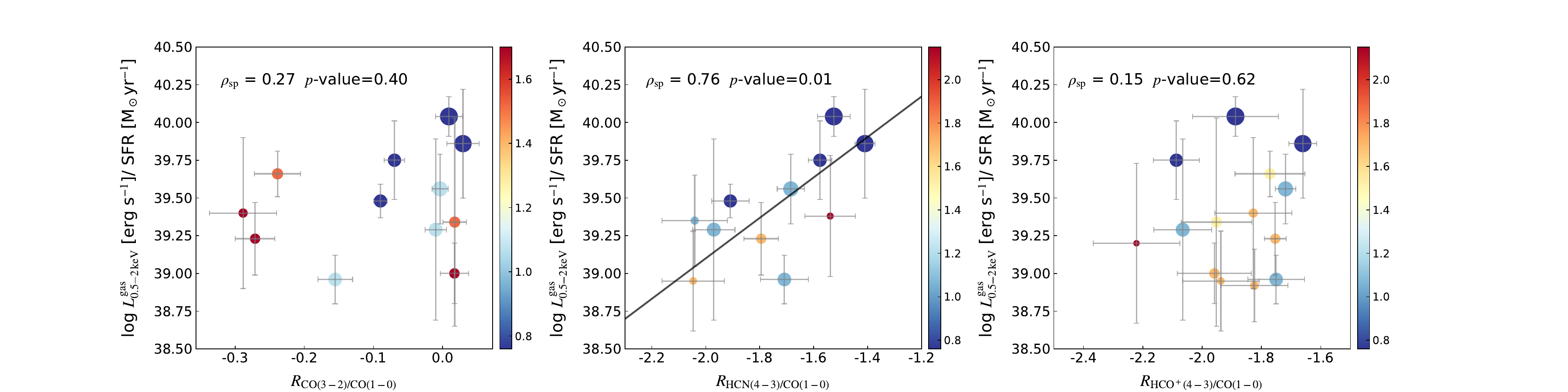}
    \caption{Relations between line ratios and and $L^{\rm gas}_{\rm 0.5-2\,keV}$/SFR on sub-kpc scale. The color scale represents the distance from the galactic center in unit of kpc, and the marker size represents the SFR surface density. The solid line in the central panel indicates the best-fit scaling relation. The Spearman rank correlation coefficient ($\rho_{\rm sp}$) and $p$-value is given in each panel. 
    }
    \label{fig:Lx-SFR-fdense}
\end{figure*}

As presented in Section \ref{Subsec:Lx-SFR}, we find a super-linear relationship between $L^{\rm gas}_{\rm\,0.5\text{–}2\,keV}$ and the SFR. The X-ray-to-SFR ratio ($L^{\rm gas}_{\rm 0.5-2\,keV}/{\rm SFR}$) indicates the relative contribution of hot gas emission compared to star formation.
To explore the possible correlation between this relative contribution and dense gas, we further investigate the relationships between $L^{\rm gas}_{\rm 0.5-2\,keV}/{\rm SFR}$ and the integrated intensity ratios CO(3–2)/CO(1–0), HCN(4–3)/CO(1–0), and HCO$^+$(4–3)/CO(1–0) (Fig. \ref{fig:Lx-SFR-fdense}). 
As the outer regions of the galaxy lack detectable dense gas, our analysis focuses solely on the central region, where HCN~(4-3) or HCO$^{+}$~(4-3) is detected with S/N $>$ 3, while excluding the central point that may be strongly influenced by the AGN.
The ratio of soft X-ray luminosity to star formation rate ($L^{\rm gas}_{\rm 0.5-2\,keV}$/SFR) represents the soft X-ray emission per unit star formation rate and serves as a valuable tracer of stellar feedback in star-forming galaxies.
We obtain log $L^{\rm gas}_{\rm 0.5-2\,keV}$/SFR = 38.5-40.3 erg~s$^{-1}$~($M_{\odot}$~yr$^{-1}$)$^{-1}$ for NGC~1068, which is consistent with the values from 
merging galaxies reported by \citet{Smith2018AJ....155...81S}.

We find no significant correlation between the X-ray-to-SFR ratio and $R_{\rm 31}$, nor between the X-ray-to-SFR luminosity ratio and $R_{\rm HCO^+(4-3)/CO(1-0)}$.  In contrast, a linear relationship is found between the X-ray-to-IR luminosity ratio and $R_{\rm HCN(4-3)/CO(1-0)}$ as follows:
\begin{align}
& \frac{\log (L^{\rm gas}_{\rm 0.5-2 keV}/\rm SFR)}{\rm erg~s^{-1}/M_{\odot}~yr^{-1}} \\
   &= 1.34(\pm0.86) \log R_{\rm HCN(4-3)/CO(1-0)}  + 41.78(\pm1.49). \nonumber
\end{align}
The correlation yields a Spearman rank coefficient of 
$\rho_{\rm sp}$ = 0.75 and a $p$-value of 0.02.
This positive correlation may arise from the relative contributions of different evolutionary stages of the star formation process, since $L^{\rm gas}_{\rm\,0.5\text{–}2\,keV}$ traces hot gas heated after star formation, SFR traced by the total infrared luminosity reflects recent or ongoing star formation \citep{Kennicutt2012ARA&A..50..531K}, HCN~(4–3) traces the dense gas directly involved in star formation, and CO~(1–0) traces the cold gas reservoir.

While both HCN and HCO$^+$ trace dense molecular gas, HCN~(4–3) has a critical density about one order of magnitude higher than that of HCO$^+$(4–3), as mentioned in Sect. \ref{Subsec:gas_IR_luminosity}. 
Given that their critical densities are relatively similar, it is unlikely that the difference in the correlation arises from the critical density itself.
The emission of HCN is strongly affected by IR radiative pumping \citep[e.g.,][]{Matsushita2015ApJ...799...26M,Stuber2023A&A...680L..20S} and XDR chemistry \citep[e.g.,][]{Lepp1996A&A...306L..21L,Meijerink2007A&A...461..793M}, whereas HCO$^+$ and CO are comparatively less impacted by these processes. 
Consequently, the HCN/CO ratio may more effectively trace the high-energy feedback, which may account for the observed super-linear correlation with $L^{\rm gas}_{\rm\,0.5\text{–}2\,keV}$/SFR.

Models and observations show that both XDR chemistry and IR radiative pumping are confined to the nuclear region near the AGN in galaxy \citep[e.g.,][]{Esposito2022MNRAS.512..686E}. Beyond this scale, their influence on molecular excitation rapidly diminishes.
Moreover, \citet{Perez-Beaupuits2007A&A...476..177P} proposed that PDRs account for the dominant excitation mechanism in the molecular gas in NGC~1068.
Note that the found super-linear correlation between  $L^{\rm gas}_{\rm\,0.5\text{–}2\,keV}$/SFR and HCN/CO  extends beyond the central region, reaching distances greater than 2 kpc.
Therefore, molecular excitation is most likely governed by AGN-related mechanisms in the nuclear region, while star formation becomes the dominant contributor in the outer regions, particularly beyond 2 kpc.

The relative enhancement of HCN emission with respect to HCO$^+$ can serve as a diagnostic of energetic processes such as AGN activity \citep{Imanishi2014AJ....148....9I}.
We found that the $R_{\rm HCN/HCO^+(4-3)}$ values range from 1 to 3, with a ratio of approximately 2 at the galaxy center. This central enhancement is significantly higher than that observed in normal or starburst galaxies, for which \citet{Tan2018ApJ...860..165T} reported an average $R_{\rm HCN/HCO^+(4-3)}$ of $0.8 \pm 0.5$ in systems without embedded AGN.
However, it is unexpected that the highest $R_{\rm HCN/HCO^+43}$ value in NGC~1068 is found off-center, reaching $\sim$3 at an offset of (0,-10). \citet{Beslic2021MNRAS.506..963B} reported that the highest HCN/HCO$^+$ ratio in NGC~3627 occurs in the bar region (approximately 2~kpc), suggesting that environments such as bars and SB rings can enhance HCN abundance. The elevated $R_{\rm HCN/HCO^+(4-3)}$ observed at $\sim$1.5~kpc in NGC~1068 may therefore indicate that the SB ring contributes to the enhancement of HCN emission, similar to the findings of \citet{Privon2015ApJ...814...39P}.

Therefore, the HCN (4-3)/CO (1-0) ratio in relation to $L^{\rm gas}_{\rm\,0.5\text{–}2\,keV}$/SFR may originate from the excitation mechanisms associated with both AGN activity and star formation.
However, the correlations are based on a small number of data points and should therefore be interpreted with caution.
To verify the correlation observed in NGC~1068 and explore its underlying mechanisms, observations with higher sensitivity and spatial resolution, together with studies extended to a larger sample, will be required.

\section{Summary} 
\label{sec:summary}
In this paper, we present the HCN $(4-3)$ and HCO$^{+}$ $(4-3)$ observational results of NGC~1068 from the MALATANG project, covering a region of approximately 1.5$^{\prime}$ $\times$ 1.5$^{\prime}$. Based on sub-scan level data reduction and incorporating spectral stacking techniques, we improved the S/N ratio in the outer disk regions. Our main conclusions are as follows:

(1) Using spectral stacking, we detected HCN~(4–3) and HCO$^+$~(4–3) emissions at approximately 3–8$\sigma$ levels in the outer disk ($\sim$ 2.2 $-$ 3.2 kpc radial distance) of NGC~1068. 

(2) The spatially resolved $L_{\rm IR} - L^{'}_{\rm dense}$
relations of NGC~1068, using HCN~(4-3) and HCO$^+ $~(4-3) as dense gas tracers, are found to be linear. 
Although the central data points could be significantly influenced by AGN activity, they do not deviate noticeably from the overall correlation.

(3) Using SED fitting, we derived a dust temperature map and found a strong positive correlation between $T_{\rm dust}$ and $R_{\rm 31}$ with a Spearman rank coefficient of $\rho_{\rm sp} = 0.60$, indicating that dust and molecular gas are thermally coupled. 
A similar positive correlation between HCN~(4–3)/CO~(1–0) and $T_{\rm dust}$ ($\rho_{\rm sp} = 0.68$), particularly in the central region of the galaxy, may be attributed to AGN feedback.
However, HCO$^+$/CO shows no correlation with $T_{\rm dust}$.

(4) A twofold correlation is found between the spatially resolved soft X-ray emission ($L^{\rm gas}_{\rm 0.5-2~keV}$) and the SFR across the galaxy, implying the presence of different feedback mechanisms linked to star formation activity. In the central region, this results in a statistically significant super-linear relationship with a slope of 2.2.

(5) A super-linear relationship is found between the soft X-ray to SFR ratio ($L^{\rm gas}_{0.5\text{–}2\,\mathrm{keV}}$/SFR) and the HCN(4–3)/CO(1–0) ratio, with a fitted slope of 1.34, which may arise from the differing contributions of various evolutionary stages in the star formation process. In contrast, this trend is not apparent for the HCO$^+$(4–3)/CO(1–0) or CO(3–2)/CO(1–0) ratios.

\section{Data availability}
Figure A1, A2, and A3 are only available in electronic form at Zenodo via url (\href{url}{https://doi.org/10.5281/zenodo.17741281}). 
Table A1 is only available in electronic form at the CDS (J/A+A/Vol/Num).

\begin{acknowledgements}
      S.L. and S.F. acknowledge support from the National Key R\&D program of China grant (2025YFE0108200), National Science Foundation of China (12373023), the starting grant at Xiamen University, and the presidential excellence fund at Xiamen University. This work is supported by the China Manned Space Program with grant no.CMS-CSST-2025-A10. Z.Y.Z acknowledges the support of NSFC (grants 12041305, 12173016), and the Program for Innovative Talents, Entrepreneur in Jiangsu.  
      Q.T. acknowledges support from the NSFC (grants 12033004, 12573013).
      Y.G. acknowledges the project ZR2024QA212 supported by Shandong Provincial Natural Science Foundation,
      National Natural Science Foundation of China (NSFC, Nos.12033004 and 12233005), and Scientific Research Fund of Dezhou University (3012304024).
      X.-L.W. acknowledges the support by the Science Foundation of Hebei Normal University (grant No. L2024B56) and S\&T Program of Hebei (grant No. 22567617H). 
      K.K. acknowledges the support by JSPS KAKENHI Grant Numbers JP17H06130, JP20H00172, JP23K20035 and JP24H00004. T.T. acknowledges the support by the NAOJ ALMA Scientific Research Grant Number 2020-15A. J.W. acknowledges the support of NSFC grants 12033004 and 12333002.

      The James Clerk Maxwell Telescope is operated by the East Asian Observatory on behalf of The National Astronomical Observatory of Japan, Academia Sinica Institute of Astronomy and Astrophysics, the Korea Astronomy and Space Science Institute, the National Astronomical Observatories of China and the Chinese Academy of Sciences, with additional funding support from the Science and Technology Facilities Council of the United Kingdom and participating universities in the United Kingdom and Canada. MALATANG is a JCMT Large Program with project codes of M16AL007 and M20AL022. We acknowledge the \texttt{ORAC-DR}, \texttt{STARLINK}, and \texttt{GILDAS} software for the data reduction and analysis. This research has made use of Chandra archival data and software provided by the Chandra X-ray Center (CXC) in the application package CIAO. This research has also made use of the NASA/IPAC Extragalactic Database (NED), which is operated by the Jet Propulsion Laboratory, California Institute of Technology, under contract with the National Aeronautics and Space Administration. The Chandra data archive is maintained by the Chandra X-ray Center at the Smithsonian Astrophysical Observatory.
      This paper is dedicated to the memory of the late Professor Yu Gao, co-PI of the MALATANG project.
\end{acknowledgements}

\bibliographystyle{aa} 
\bibliography{ref}

\begin{appendix}

\section{Intensities and luminosities table}
\label{Appendix: C}
The intensities and luminosities of different tracers from individual resolved regions are provided in electronic form at the CDS (J/A+A/Vol/Num).

\section{SED fitting results}
\label{Appendix: D}
\begin{figure*}
    \centering
    \includegraphics[width=\linewidth]{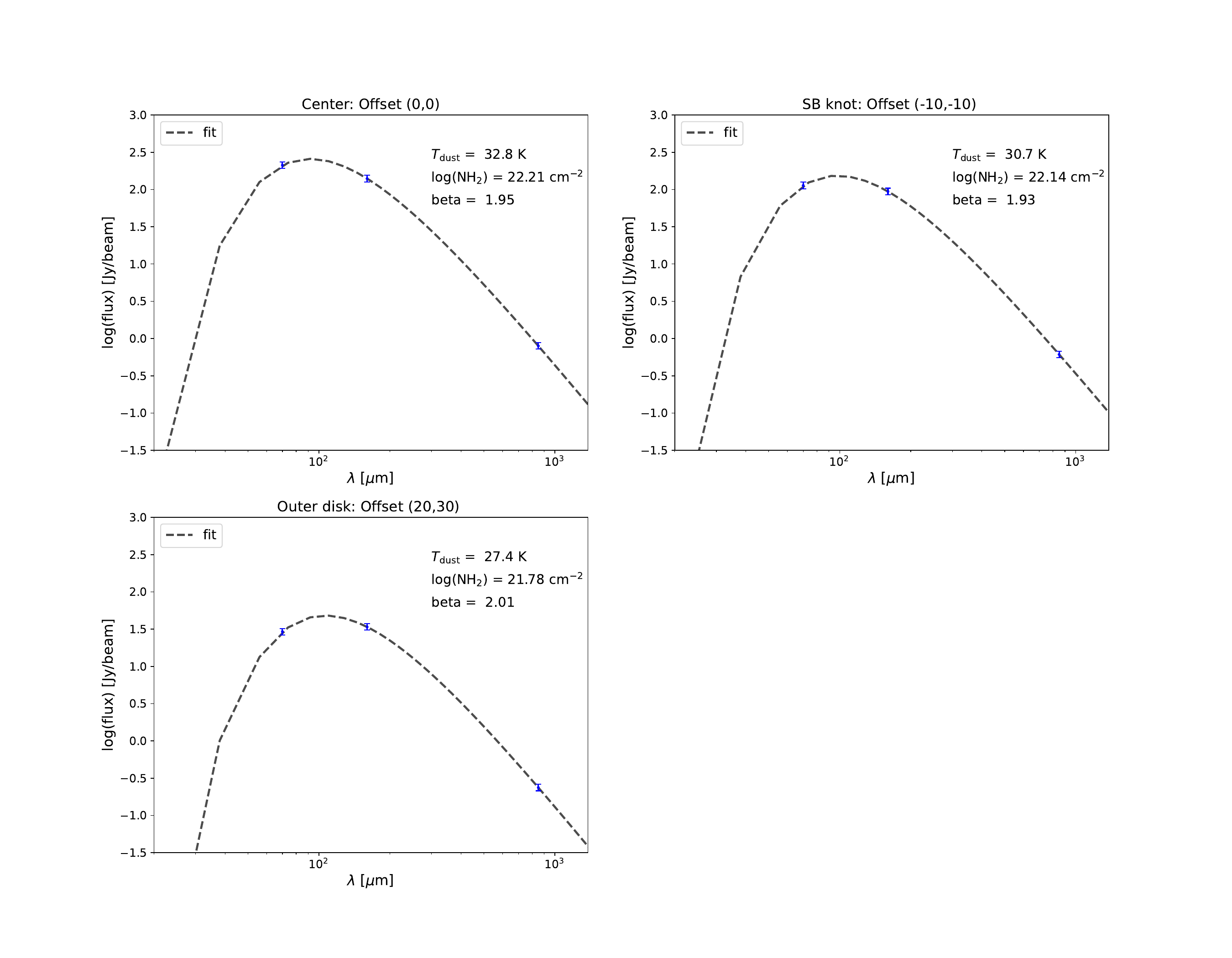}
    \caption{Dust SEDs of NGC 1068 at three representative positions: the galaxy center (0,0), a starburst knot (-10,-10), and the outer disk (20,30). Blue dots represent the photometric data at 70~$\mu$m, 160~$\mu$m, and 850~$\mu$m, and the dashed lines represent the best-fit modified blackbody model. The fitting results are shown in the upper-left corner of each panel.}
    \label{fig:SED_fit}
\end{figure*}

Figure \ref{fig:SED_fit} presents the dust SEDs of NGC~1068 at three representative positions: the galaxy center (0,0), a starburst knot (-10, -10), and the outer disk (20, 30).

\FloatBarrier 
\twocolumn

\end{appendix}

\end{document}